\documentclass[aps,pra,reprint,superscriptaddress]{revtex4-2}
\usepackage{amsmath,amssymb,bm}
\usepackage{dsfont}
\usepackage{graphicx}
\graphicspath{{figures/}}
\usepackage{xcolor}
\usepackage[colorlinks=true,allcolors=blue]{hyperref}
\usepackage{orcidlink}

\makeatletter
\newcounter{outlineitem}

\makeatother
\newcommand{\sx}{\sigma^{x}}

\newcommand{\sz}{\sigma^{z}}

\newcommand{\DTWA}{\mathrm{DTWA}}

\begin{document}

\title{Discrete truncated Wigner approximation of the quantum phase transition in the transverse-field Ising model}

\author{Tuan-Vu Truong\,\orcidlink{0009-0005-6357-6060}}
\affiliation{Identity Quantum Computing JSC, 117 Tran Duy Hung Street, Hanoi 10000, Vietnam}
\affiliation{College of Natural Sciences, Can Tho University, Can Tho, Vietnam}
\author{Hoang-Long Nguyen\,\orcidlink{0009-0005-7922-9266}}
\affiliation{Identity Quantum Computing JSC, 117 Tran Duy Hung Street, Hanoi 10000, Vietnam}
\date{\today}

\begin{abstract}
The discrete truncated Wigner approximation (DTWA) has been a workhorse for large-scale spin dynamics, but whether it captures dynamics across a quantum phase transition has not been tested systematically.
In this work, we use the transverse-field Ising model (TFIM) to study the quantum phase transition with DTWA from one to three dimensions and with all-to-all connectivity, using slow linear ramps and fast ramps.
We find that DTWA cannot capture the dynamics on the one-dimensional chain, and that its behavior is consistent with classical thermalization, in which its classical spins relax toward a thermal state at a finite effective temperature that does not vanish for slow ramps, while a one-dimensional chain cannot order at any finite temperature.
For square and cubic lattices, in contrast to the chain, DTWA builds order, and the order sets in close to the quantum critical point found by quantum Monte Carlo (QMC).
However, its correlations do not grow with the Kibble--Zurek exponents, so DTWA captures the onset of order but not the universal dynamics of the transition.
For all-to-all coupling, DTWA follows the exact growth of the order parameter along the entire ramp.
Our results suggest that DTWA becomes more reliable as each spin interacts with more partners, and that beyond one dimension it can indicate where a quantum phase transition lies but cannot capture its critical scaling.
\end{abstract}

\maketitle

\section{Introduction}
Quantum simulators built from Rydberg atoms~\cite{Bernien2017,Keesling2019,Ebadi2021} and trapped ions~\cite{Zhang2017}, as well as quantum annealers~\cite{King2022,King2023}, now prepare hundreds or more interacting spins and drive them across quantum phase transitions, for example in Kibble--Zurek experiments~\cite{Keesling2019,King2022}.
Simulating such dynamics classically is hard beyond one dimension, because exact methods are limited to small systems and tensor-network and neural-network methods remain costly in two dimensions~\cite{Schmitt2022}.
Due to this constraint, cheap approximate methods that scale to large two- and three-dimensional systems are essential.

To handle this constraint in computational capacity, one candidate, the truncated Wigner approximation~\cite{Steel1998,Polkovnikov2010,Davidson2015}, replaces quantum dynamics with classical trajectories whose random initial conditions encode the quantum fluctuations of the initial state.
Its discrete version for spin one-half, DTWA~\cite{Schachenmayer2015,Schachenmayer2015NJP}, samples these initial conditions from a discrete Wigner function~\cite{Wootters1987,Gibbons2004} and has proven to be an effective method for simulating large spin systems~\cite{Zhu2019}.
It works well when each spin interacts with many partners, for long-range interactions and short times~\cite{Kunimi2021}, and it becomes exact in the mean-field limit~\cite{Polkovnikov2010}.

However, DTWA becomes less reliable for short-range interactions.
In one-dimensional quenches near criticality, it misses long-distance correlations and its correlation length saturates at the wrong value~\cite{Czischek2018}.
In two dimensions, it describes dynamical transitions only qualitatively~\cite{Khasseh2020}, and for short-range couplings it departs from tensor-network results at long times~\cite{Muleady2023}.
Classical-field methods for bosons are also known to relax to classical thermal equilibrium~\cite{Sinatra2002,Blakie2008}, an artifact that turns out to be essential in this work.
Thermalization has also been studied directly with DTWA in disordered spin chains~\cite{Acevedo2017}.
Attempts to improve DTWA, such as adding two-spin correlations~\cite{Pucci2016}, which are unstable at late times for this model~\cite{Czischek2018}, or treating clusters of spins exactly~\cite{Wurtz2018,Wurtz2020}, mainly improve short distances.

The question of whether DTWA captures dynamics across a quantum phase transition, and how its accuracy depends on dimension, has not been tested systematically.
Slow ramps through the critical point are a demanding test for three reasons.
First, the Kibble--Zurek mechanism predicts universal scaling with known exponents~\cite{Kibble1976,Zurek1985,ZurekDornerZoller2005,Dziarmaga2005,Damski2005,Polkovnikov2005,Dziarmaga2010,Polkovnikov2011,delCampo2014}.
Second, slow ramps reach long times where classical artifacts accumulate.
Third, benchmarks exist against which DTWA can be checked, namely the exact free-fermion solution in one dimension~\cite{Pfeuty1970}, the exact dynamics of the all-to-all (Lipkin--Meshkov--Glick) model~\cite{LMG1965}, quantum Monte Carlo (QMC) critical points~\cite{BloteDeng2002} and near-exact Kibble--Zurek results in two dimensions~\cite{Schmitt2022}.

In this work, we ramp the transverse-field Ising model (TFIM) across its quantum phase transition on the one-dimensional chain, on square and cubic lattices and with all-to-all coupling, using slow linear ramps and fast ramps.
On the chain, DTWA cannot capture the growth of the correlations, and its state is consistent with classical spins that thermalize at a finite effective temperature that slow ramps do not remove, while a one-dimensional chain cannot order at any finite temperature.
The same thermal state accounts for the nearest-neighbor correlator exceeding the quantum bound.
Because no quantum state can exceed this bound, the violation is a simple warning sign that a DTWA simulation has become unreliable, although its absence does not guarantee accuracy.
On square and cubic lattices, the spins thermalize in the same way, but a classical magnet can order at a finite temperature in two and three dimensions.
DTWA therefore builds order, which sets in close to the QMC critical point, but its correlations grow by classical relaxation rather than by the quantum Kibble--Zurek mechanism, so they do not follow the Kibble--Zurek exponents.
With all-to-all coupling, DTWA follows the exact dynamics quantitatively.
Treating small clusters of spins exactly improves the nearest-neighbor correlator but not the long-distance correlations.

From this study, we find that DTWA becomes more reliable as each spin interacts with more partners, and that the classical thermalization of its spins is consistent with its behavior in all three dimensions.
Beyond one dimension, DTWA can therefore indicate where a quantum phase transition lies but cannot capture its critical scaling.
The paper is organized as follows.
Section~\ref{sec:model} introduces the model and the ramp protocols.
Section~\ref{sec:methods} describes DTWA and cluster TWA.
Sections~\ref{sec:1d}, \ref{sec:lmg} and~\ref{sec:2d3d} present the results for the chain, for all-to-all coupling and for square and cubic lattices.
Section~\ref{sec:mechanism} shows that these behaviors are consistent with classical thermalization in all three dimensions.
Section~\ref{sec:discussion} discusses the implications.

\section{Model and protocols}
\label{sec:model}
The transverse-field Ising model is expressed by the following Hamiltonian,
\begin{equation}
  H = -J\sum_{\langle ij\rangle}\sz_i\sz_j \;-\; h\sum_i \sx_i ,
  \label{eq:tfim}
\end{equation}
with ferromagnetic nearest-neighbor coupling $J$ and transverse field $h$, on a periodic chain of $N$ sites, on square lattices of $L\times L$ sites and on simple cubic lattices of $L^3$ sites, all with periodic boundary conditions.
Its ground state changes from a paramagnet polarized along $x$ at large $h$ to a ferromagnet ordered along $z$ at large $J$, through a quantum critical point.
We use units with $\hbar=1$ throughout.

We also consider an all-to-all version of the model, in which every spin couples equally to all others with strength $J/N$.
Its Hamiltonian is
\begin{equation}
\begin{split}
  H &= -h\sum_i \sx_i - \frac{J}{N}\sum_{i<j}\sz_i\sz_j \\
    &= -2hS^x - \frac{J}{N}\left(2(S^z)^2-\frac{N}{2}\right).
\end{split}
  \label{eq:lmg}
\end{equation}
Because this Hamiltonian depends only on the collective spin $\mathbf S=\frac12\sum_i\bm\sigma_i$, the dynamics stays in the $(N+1)$-dimensional symmetric (Dicke) subspace and can be solved exactly for large $N$.
The model has a ferromagnetic transition at $J_c=h$.

The all-to-all model is the mean-field limit of the TFIM.
In this limit, TWA is expected to become exact, with corrections of order $1/N$~\cite{Polkovnikov2010}.
Below, we define the two ramp protocols, the observables and the Kibble--Zurek predictions against which DTWA is tested.

\subsection{Linear interpolation ramp}
Our main protocol interpolates linearly between the paramagnetic and the ferromagnetic Hamiltonians.
The ramp is expressed by the time-dependent Hamiltonian
\begin{equation}
  H(s) = -(1-s)\sum_i \sx_i \;-\; s\sum_{\langle ij\rangle}\sz_i\sz_j ,
  \label{eq:ramp}
\end{equation}
with $s=t/T$ running from 0 to 1 over the ramp time $T$.
The ramp crosses the critical point at
\begin{equation}
  s_c = \frac{1}{1+h_c/J}.
  \label{eq:sc}
\end{equation}
For all-to-all coupling, the nearest-neighbor sum in Eq.~\eqref{eq:ramp} is replaced by the all-to-all coupling of Eq.~\eqref{eq:lmg},
\begin{equation}
  \sum_{\langle ij\rangle}\sz_i\sz_j\;\longrightarrow\;\frac{1}{N}\sum_{i<j}\sz_i\sz_j .
  \label{eq:a2a}
\end{equation}
We initialize all $N$ spins along $+x$, in the product state $|{+}x\rangle^{\otimes N}$, which is the exact ground state at $s=0$, and end the ramp at $s=1$, where $h=0$.

\subsection{Fast-ramp protocol}
To compare directly with near-exact results in two dimensions, we also use the fast-ramp protocol of Ref.~\cite{Schmitt2022},
\begin{equation}
\begin{gathered}
  \frac{h(t)}{h_c}=1-\frac{t}{\tau_Q},\qquad \frac{J(t)}{J_c}=1+\frac{t}{\tau_Q},\\
  t\in[-\tau_Q,0].
\end{gathered}
\label{eq:schmitt}
\end{equation}
The field and the coupling are both ramped linearly in time, so that the critical point is reached at $t=0$, and the quench time $\tau_Q$ sets how fast it is approached.
Here $h_c/J_c$ is the critical ratio of the lattice, which is one on the chain and given by the QMC values of Sec.~\ref{sec:2d3d} on the square and cubic lattices, $J_c$ sets the unit of energy, and $\tau_Q$ is measured in units of $1/J_c$.
At $t=0$ we record the equal-time correlation function
\begin{equation}
  C(r)=\langle\sz_i\sz_{i+r}\rangle ,
  \label{eq:czz}
\end{equation}
between two spins a distance $r$ apart along a lattice axis, averaged over all sites and lattice axes.
If the dynamics follows the Kibble--Zurek mechanism, the correlations at $t=0$ obey the scaling form
\begin{equation}
\begin{gathered}
  \hat\xi^{\,2\Delta}\,C(r)\big|_{t=0}=F_C\!\left(r/\hat\xi\right),\\
  \hat\xi=\tau_Q^{\,a},\qquad 2\Delta=1+\eta,\quad \eta=0.0363 ,
\end{gathered}
\label{eq:collapse}
\end{equation}
where $2\Delta$ is the scaling dimension of $\sz_i$ on the square lattice and $\eta$ is the anomalous exponent of the classical three-dimensional Ising model; on the chain, $2\Delta=1/4$. The curves for different quench times then collapse onto a single curve $F_C$ when $a$ takes its Kibble--Zurek value.
For this protocol, Ref.~\cite{Schmitt2022} found a Kibble--Zurek exponent of $0.36$ on the square lattice, slightly below the asymptotic value of Table~\ref{tab:kz}, which serves as our two-dimensional benchmark.

\subsection{Observables}
We characterize the state by the equal-time correlation function $C(r)$ of Eq.~\eqref{eq:czz}, together with four quantities derived from it.
In DTWA, all of them are evaluated with the estimators of Sec.~\ref{sec:methods}.

The order parameter is expressed as
\begin{equation}
  m_z^2=\frac{1}{N(N-1)}\sum_{i\neq j}\langle\sz_i\sz_j\rangle ,
  \label{eq:m2}
\end{equation}
the average correlation between all pairs of distinct spins.
It vanishes in the paramagnet and approaches one in a fully ordered ferromagnet.

The correlation length $\xi_{1/e}$ is defined by
\begin{equation}
  C(\xi_{1/e})=e^{-1} ,
  \label{eq:xi}
\end{equation}
the distance at which the correlation function drops to $e^{-1}$, with $\ln C(r)$ interpolated linearly between lattice sites.

The kink density at the end of the ramp is
\begin{equation}
  n_{\rm kink}=\tfrac12\left(1-\langle\sz_i\sz_{i+1}\rangle\right)\Big|_{s=1} ,
  \label{eq:nkink}
\end{equation}
the fraction of nearest-neighbor bonds whose two spins point in opposite directions, that is, the density of domain walls left behind by the ramp.

Finally, we locate the transition by using the Binder cumulant
\begin{equation}
  U=1-\frac{\langle M^4\rangle}{3\langle M^2\rangle^2},\qquad M=\frac1N\sum_i s^z_i ,
  \label{eq:binder}
\end{equation}
where $M$ is the magnetization of a single DTWA trajectory, built from the classical spin components $s^z_i$ of Sec.~\ref{sec:methods}, so that the cumulant measures where the classical ensemble of trajectories orders.
It vanishes for Gaussian fluctuations of $M$ in the disordered phase and approaches $2/3$ in the ordered phase.

\subsection{Kibble--Zurek expectations}
For a slow ramp, the Kibble--Zurek mechanism describes how the system falls out of equilibrium near the critical point~\cite{Kibble1976,Zurek1985,ZurekDornerZoller2005,Dziarmaga2005}.
Far from the critical point the energy gap is large, and the state follows the instantaneous ground state.
Close to it the gap closes and the relaxation time diverges, so the state can no longer follow the ramp and its correlations freeze.
Balancing the relaxation time against the time left before the critical point gives the frozen correlation length
\begin{equation}
  \hat\xi\propto T^{\,a},\qquad a=\frac{\nu}{1+z\nu},
  \label{eq:kz}
\end{equation}
where $\nu$ is the correlation-length exponent and $z$ the dynamic exponent of the transition.
Slower ramps therefore leave larger correlated regions behind.

On the chain, roughly one kink is left behind per correlated region, so the kink density scales as
\begin{equation}
  n_{\rm kink}\propto\hat\xi^{\,-1}\propto T^{-a}.
  \label{eq:kzkink}
\end{equation}

For the TFIM the dynamic exponent is $z=1$, so that $a=\nu/(1+\nu)$.
The quantum model in $d$ dimensions belongs to the universality class of the classical Ising model in $d+1$ dimensions~\cite{Suzuki1976,Sachdev2011}, which gives the exponents listed in Table~\ref{tab:kz}.
In three dimensions the model is at its upper critical dimension, so the mean-field value $\nu=1/2$ applies up to logarithmic corrections.

\begin{table}
  \caption{Kibble--Zurek exponents expected for the TFIM ($z=1$). The quantum model in $d$ dimensions belongs to the universality class of the classical Ising model in $d+1$ dimensions.}
  \label{tab:kz}
  \begin{ruledtabular}
  \begin{tabular}{lccc}
   & universality & $\nu$ & $a=\nu/(1+\nu)$ \\ \hline
   1D & 2D Ising & 1 & 1/2 \\
   2D & 3D Ising & 0.630 & 0.386 \\
   3D & 4D Ising (MF, logs) & 1/2 & 1/3 \\
  \end{tabular}
  \end{ruledtabular}
\end{table}

\section{Methods}
\label{sec:methods}
In this section, we explain how we simulate the ramps with DTWA and, on the chain, also with a cluster extension of it.
We compare both with exact or near-exact references, which are the free-fermion solution on the chain~\cite{Pfeuty1970}, the exact solution in the Dicke subspace for all-to-all coupling, QMC critical points on square and cubic lattices~\cite{BloteDeng2002}, and near-exact Kibble--Zurek results on the square lattice~\cite{Schmitt2022}, which are described in App.~\ref{app:exact}.

\subsection{DTWA}
DTWA describes each spin one-half by a classical vector $\mathbf s_i$ on a discrete phase space~\cite{Schachenmayer2015}.
This phase space consists of the eight points $\mathbf s=(s^x,s^y,s^z)$ with every component equal to $\pm1$, the corners of a cube.
Each phase point corresponds to a phase-point operator $\hat A_{\mathbf s}$, and the discrete Wigner function of a single-spin density matrix $\rho$ is the expectation value of these operators,
\begin{equation}
\begin{gathered}
  W(\mathbf s)=\tfrac12\big\langle\hat A_{\mathbf s}\big\rangle=\tfrac12\,\mathrm{Tr}\big[\rho\,\hat A_{\mathbf s}\big],\\
  \hat A_{\mathbf s}=\tfrac12\big(\mathds{1}+\mathbf s\cdot\bm\sigma\big),\qquad
  \mathbf s\in\{-1,+1\}^3 .
\end{gathered}
\label{eq:wigner}
\end{equation}
The eight points split into two sets of four, $\Gamma_+$ and $\Gamma_-$, which contain the points with $s^xs^ys^z=+1$ and $s^xs^ys^z=-1$, respectively.
Within each set, $W$ is normalized and reproduces all single-spin expectation values,
\begin{equation}
  \sum_{\mathbf s\in\Gamma_\pm}W(\mathbf s)=1,\qquad
  \sum_{\mathbf s\in\Gamma_\pm}W(\mathbf s)\,s^a=\langle\sigma^a\rangle ,
  \label{eq:wigner_norm}
\end{equation}
for $a=x,y,z$.
For the product state $|{+}x\rangle^{\otimes N}$, the many-spin Wigner function is a product of single-spin ones.
Each spin has $W=1/2$ on the two points of a set with $s^x=1$, and zero elsewhere.
Averaging over both sets~\cite{Schachenmayer2015} gives the following sampling rule,
\begin{equation}
  s^x_i=1,\qquad s^y_i,\,s^z_i\in\{-1,+1\},
  \label{eq:sampling}
\end{equation}
with each sign of $s^y_i$ and $s^z_i$ drawn independently with probability one half.
Every sampled vector has length $|\mathbf s_i|=\sqrt3$, the classical counterpart of $\bm\sigma^2=3$.
The dynamics is generated by the Weyl symbol $H_{\rm W}$ of the Hamiltonian.
For an operator $\hat O$, the Weyl symbol is the phase-space function
\begin{equation}
  O_{\rm W}(\mathbf s)=\mathrm{Tr}\big[\hat O\,\hat A_{\mathbf s}\big],\qquad
  \langle\hat O\rangle=\sum_{\mathbf s\in\Gamma_\pm}W(\mathbf s)\,O_{\rm W}(\mathbf s),
  \label{eq:weyl}
\end{equation}
which turns expectation values into averages over the phase space.
For a Pauli operator it is $(\sigma^a)_{\rm W}=s^a$, and for a product of operators on different sites it is the product of their symbols.
The Weyl symbol of the Hamiltonian is therefore obtained by replacing each Pauli operator by the corresponding component of $\mathbf s_i$, $\sigma^a_i\to s^a_i$,
\begin{equation}
  H_{\rm W}=-J\sum_{\langle ij\rangle}s^z_is^z_j-h\sum_i s^x_i .
  \label{eq:hw}
\end{equation}
This replacement is exact here, because the Hamiltonian contains only single-spin terms and products of operators on different sites.
Next, we consider how the Wigner function evolves in time. For a system with canonical phase-space variables $(q,p)$, it evolves exactly according to the Moyal bracket~\cite{Polkovnikov2010},
\begin{equation}
\begin{split}
  \frac{\partial W}{\partial t}&=\frac{2}{\hbar}\,H_{\rm W}\sin\!\Big(\frac{\hbar}{2}\Lambda\Big)W\\
  &=\{H_{\rm W},W\}-\frac{\hbar^2}{24}\,H_{\rm W}\Lambda^3W+O(\hbar^4),
\end{split}
\label{eq:moyal}
\end{equation}
where $\hbar$, set to one elsewhere, is kept here to show the order of each term, and $\Lambda=\overleftarrow{\partial_q}\overrightarrow{\partial_p}-\overleftarrow{\partial_p}\overrightarrow{\partial_q}$ acts to the left on $H_{\rm W}$ and to the right on $W$, so that $H_{\rm W}\Lambda W=\{H_{\rm W},W\}$ is the Poisson bracket. The first term alone is the classical Liouville equation. The remaining terms are quantum corrections of order $\hbar^2$ and higher, with three or more derivatives. Since they vanish when $H_{\rm W}$ is at most quadratic in $q$ and $p$ and are small when the dynamics is close to classical, they can be truncated.
For spins, the truncation is exact when the Hamiltonian is linear in the spin operators, and it becomes accurate when each spin feels the average field of many partners.
After the truncation, the Wigner function obeys the classical Liouville equation
\begin{equation}
  \frac{\partial W}{\partial t}=\{H_{\rm W},W\},
  \label{eq:liouville}
\end{equation}
which is solved by moving every sample along its classical trajectory, with $W$ constant along each trajectory.
For spins, the discrete phase space has no exact Moyal equation of this form, so evolving the discrete samples classically is an ansatz, which the comparisons with exact results below test.

To write these trajectories for spins, we need the classical counterpart of the spin commutation relations.
Here, the classical spins obey the Poisson bracket
\begin{equation}
  \{s^a_i,s^b_j\}=2\,\delta_{ij}\,\epsilon_{abc}\,s^c_i ,
  \label{eq:poisson}
\end{equation}
the classical image of the commutator $[\sigma^a_i,\sigma^b_j]=2i\,\delta_{ij}\,\epsilon_{abc}\,\sigma^c_i$, where $\epsilon_{abc}$ is the Levi-Civita symbol.
For two functions $A$ and $B$ of the spins, this bracket reads
\begin{equation}
  \{A,B\}=2\sum_i\epsilon_{abc}\,s^c_i\,\frac{\partial A}{\partial s^a_i}\frac{\partial B}{\partial s^b_i},
  \label{eq:spinbracket}
\end{equation}
with summation over the repeated component indices $a,b,c$. It reduces to Eq.~\eqref{eq:poisson} for $A=s^a_i$ and $B=s^b_j$, and it is the bracket that enters the Liouville equation~\eqref{eq:liouville} for spins.
Hamilton's equations, $d\mathbf s_i/dt=\{\mathbf s_i,H_{\rm W}\}$, then give
\begin{equation}
  \frac{ds^a_i}{dt}=\{s^a_i,H_{\rm W}\}=2\,\epsilon_{abc}\,s^c_i\,\frac{\partial H_{\rm W}}{\partial s^b_i},
  \label{eq:hamilton}
\end{equation}
which in vector form is the precession
\begin{equation}
\begin{gathered}
  \frac{d\mathbf{s}_i}{dt}=2\,\mathbf{s}_i\times\mathbf{B}_i,\\
  \mathbf{B}_i=-\frac{\partial H_{\rm W}}{\partial \mathbf{s}_i}=\Big(h,\,0,\,J\sum_{j\in\partial i}s^z_j\Big),
\end{gathered}
\label{eq:eom}
\end{equation}
where $\mathbf B_i$ is the local field acting on spin $i$. Its $x$ component is the transverse field $h$, and its $z$ component is the Ising field $J\sum_{j\in\partial i}s^z_j$ produced by the nearest neighbors $\partial i$ of site $i$. For all-to-all coupling, the sum runs over all other spins with weight $1/N$. Each spin therefore precesses about its local field at angular frequency $2|\mathbf B_i|$, which conserves its length $|\mathbf s_i|$, while the field itself changes as the neighbors move.

Since Eq.~\eqref{eq:eom} has no closed-form solution once the spins interact, we integrate it numerically in time steps $\delta t$.
Within each step, we split the local field into its transverse part, which rotates every spin about the $x$ axis, and its Ising part, which rotates every spin about the $z$ axis.
The transverse part is a rotation about the fixed $x$ axis at the rate $2h$. The Ising part is a rotation about the $z$ axis at the rate $2J\sum_{j\in\partial i}s^z_j$, and this rate stays constant during the step, because a rotation about $z$ does not change any $s^z_j$. Both parts can therefore be carried out exactly.
We then apply them in the symmetric order of a half step of the transverse part, a full step of the Ising part and another half step of the transverse part, which is accurate to second order in $\delta t$ and conserves every length $|\mathbf s_i|$ exactly.
In operator form, one time step reads
\begin{equation}
  e^{\delta t\,\mathcal L}\approx e^{\frac{\delta t}{2}\mathcal L_X}\,e^{\delta t\,\mathcal L_Z}\,e^{\frac{\delta t}{2}\mathcal L_X},
  \label{eq:strang}
\end{equation}
where $\mathcal L=\mathcal L_X+\mathcal L_Z$ generates the motion of Eq.~\eqref{eq:eom}, with $\mathcal L_X$ the rotation by the transverse field and $\mathcal L_Z$ the rotation by the Ising field.

With the samples evolved, we can now measure observables. In DTWA, the expectation value of an operator at time $t$ is the average of its Weyl symbol, Eq.~\eqref{eq:weyl}, over the evolved samples, $\langle\hat O(t)\rangle\approx\overline{O_{\rm W}\big(\mathbf s(t)\big)}$, where the overline denotes the average over trajectories. For the correlations and collective moments used in this work, this gives
\begin{align}
  \langle \sz_i\sz_j\rangle_{\DTWA} &= \overline{s^z_is^z_j}\quad (i\neq j), \label{eq:est1}\\
  \langle (S^a)^2\rangle_{\DTWA} &= \frac N4+\frac14\sum_{k\neq l}\overline{s^a_ks^a_l}. \label{eq:est2}
\end{align}
The first expression states that the correlation between two different sites is the average of the product of their classical components, because the Weyl symbol of a product of operators on different sites is the product of their symbols. Since this average runs over a finite number of trajectories, it has a statistical noise floor at large distances, and we therefore use the $1/e$ definition of the correlation length, Eq.~\eqref{eq:xi}, rather than the second moment of $C(r)$, which is dominated by this noisy tail. The second expression follows from writing $(S^a)^2=\frac14\sum_{k,l}\sigma^a_k\sigma^a_l$. The terms with $k=l$ give the self-term $N/4$, because $(\sigma^a_k)^2=1$ for every spin, while the terms with $k\neq l$ are again averages of products.

\subsection{Gaussian cluster TWA}
To test whether a better semiclassical method repairs the chain, we also use the Gaussian cluster TWA~\cite{Wurtz2018}.
The chain is divided into clusters of $m$ neighboring sites.
For $m=1$ every spin forms its own cluster and the method reduces to the Gaussian TWA of single spins, while for $m=N$ it is exact, so increasing $m$ makes the description progressively more quantum.
Each cluster $c$ is described by a $2^m$-dimensional matrix $M_c(t)$, which plays the role of the cluster's density matrix. The classical value of any operator $\hat O$ acting on the cluster follows from it as
\begin{equation}
  O_c=\mathrm{Tr}\big[M_c\,\hat O\big],
  \label{eq:ctwa_M}
\end{equation}
and, unlike a density matrix, $M_c$ need not be positive.
Quantum fluctuations enter through the initial value of $M_c$, as they do through the random signs of Eq.~\eqref{eq:sampling} in DTWA. Instead of starting from the density matrix $|\psi\rangle\langle\psi|$ of the initial cluster state $|\psi\rangle=|{+}x\rangle^{\otimes m}$, each sample starts from
\begin{equation}
\begin{gathered}
  M_c(0) = |\psi\rangle\langle\psi| + |\psi\rangle\langle\chi| + |\chi\rangle\langle\psi|,\\
  |\chi\rangle=\tfrac{1}{\sqrt2}\big(|w\rangle-|\psi\rangle\langle\psi|w\rangle\big),
\end{gathered}
\label{eq:ctwa_sample}
\end{equation}
where $|w\rangle$ is a vector of $2^m$ independent complex Gaussian random numbers $w_k$ with zero mean and unit variance. Removing its component along $|\psi\rangle$ makes $|\chi\rangle$ a random vector orthogonal to $|\psi\rangle$. Averaged over the samples, $M_c(0)$ equals $|\psi\rangle\langle\psi|$, so all mean values are those of the initial state, while the fluctuations between samples reproduce the symmetrized quantum covariances of all generators of $\mathfrak{su}(2^m)$~\cite{Wurtz2018}.

The matrix $M_c$ then evolves exactly under the cluster Hamiltonian $H_c$, the TFIM restricted to the $m$ sites of the cluster,
\begin{equation}
  H_c=-J\sum_{k=1}^{m-1}\sz_k\sz_{k+1}-h\sum_{k=1}^{m}\sx_k ,
  \label{eq:ctwa_hc}
\end{equation}
while neighboring clusters interact through their mean fields,
\begin{equation}
  H^{\rm MF}_c=-J\Big[\mathrm{Tr}\big(M_{c-1}\sz_m\big)\,\sz_1+\mathrm{Tr}\big(M_{c+1}\sz_1\big)\,\sz_m\Big],
  \label{eq:ctwa_mf}
\end{equation}
in which the edge spins of cluster $c$ feel the classical $z$ values of the adjacent edge spins of the neighboring clusters. Together, the evolution of each cluster is
\begin{equation}
  \dot M_c=-i\big[H_c+H^{\rm MF}_c,\,M_c\big].
  \label{eq:ctwa_eom}
\end{equation}

As in DTWA, observables are averages over the samples. For the spin correlations, this gives
\begin{align}
  \langle\sz_i\sz_j\rangle &= \overline{\mathrm{Tr}\big[M_c\,\sz_i\sz_j\big]}, \label{eq:ctwa_meas1}\\
  \langle\sz_i\sz_j\rangle &= \overline{\mathrm{Tr}\big[M_c\,\sz_i\big]\,\mathrm{Tr}\big[M_{c'}\sz_j\big]}. \label{eq:ctwa_meas2}
\end{align}
The first expression holds for two spins in the same cluster $c$ and follows directly from Eq.~\eqref{eq:ctwa_M}. The second holds for spins in different clusters $c$ and $c'$, whose correlation is the average of a product of single-cluster values, as for different sites in DTWA.

\section{One dimension}
\label{sec:1d}
\begin{figure}[!t]
  \centering
  \includegraphics[width=0.88\columnwidth]{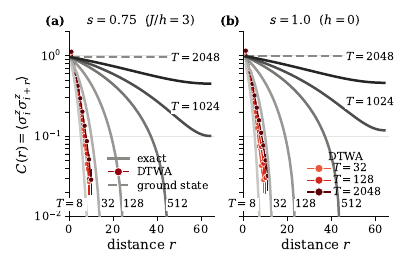}
  \caption{Spin correlations on the chain during the linear ramp, (a) at $s=0.75$ ($J/h=3$) and (b) at the end of the ramp, $s=1$ ($h=0$), for $N=128$. Gray lines show the exact solution, labeled by the ramp time $T$ and shaded from light (fast) to dark (slow), and red points show DTWA for $T=32$, 128 and 2048, shaded in the same way from light red (fast) to dark red (slow). DTWA points consistent with zero within two standard errors are omitted. The dashed line is the exact ground state. The exact correlations spread to longer distances as the ramp slows, whereas the DTWA correlations grow only slightly and remain short-ranged.}
  \label{fig:1d_corr}
\end{figure}
\begin{figure}[!t]
  \centering
  \includegraphics[width=0.88\columnwidth]{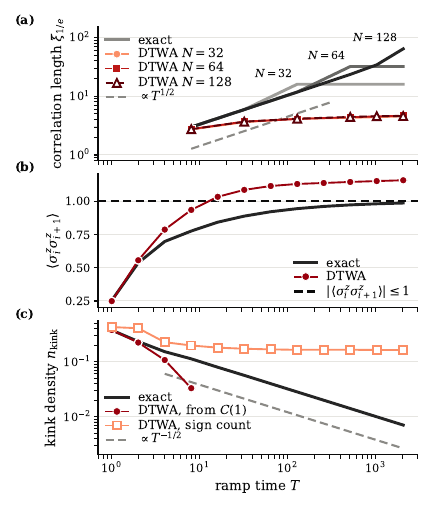}
  \caption{The chain at the end of the ramp, $s=1$ ($h=0$), versus ramp time $T$. (a) Correlation length $\xi_{1/e}$ for $N=32$, 64 and 128. Gray lines show the exact solution, shaded from light (small $N$) to dark (large $N$), and red symbols show DTWA, with $N=128$ drawn dashed with open symbols so that the curves beneath it remain visible. The exact correlation length saturates at $N/2$ once the whole ring is correlated, whereas the DTWA curves for all three system sizes coincide. (b) Nearest-neighbor correlator, exact for $N=1024$ and DTWA for $N=256$, with the quantum bound $|\langle\sz_i\sz_{i+1}\rangle|\le 1$ dashed. (c) Kink density, exact for $N=1024$ (black line) and DTWA for $N=256$, estimated from the nearest-neighbor correlator through Eq.~\eqref{eq:nkink} (filled circles, shown only while positive) and by counting sign changes of $s^z_i$ between neighbors (open squares). The dashed lines in (a) and (c) show the Kibble--Zurek power laws $T^{1/2}$ and $T^{-1/2}$.}
\label{fig:fig2-correlation-length}
\end{figure}
In this section, we study ramps across the quantum critical point of the one-dimensional chain.
The chain is the most demanding test for a semiclassical method, because quantum fluctuations are strongest in one dimension, and it is also the case in which the exact dynamics of large systems is available from the free-fermion solution~\cite{Pfeuty1970}.
The critical point lies at $h_c=J$, so the linear ramp of Eq.~\eqref{eq:ramp} crosses it halfway, at $s_c=1/2$.
We compare DTWA with the exact solution after the ramp has crossed the critical point, mainly at its end, where $h=0$.
At the critical point itself the correlations on the linear ramp are still short, so we test it separately with the fast-ramp protocol of Eq.~\eqref{eq:schmitt}.
We first compare the correlation functions and the correlation length, then turn to the nearest-neighbor correlator and the kink density, and finally ask whether treating clusters of spins exactly improves the agreement.

We first compare the correlation function $C(r)$ of Eq.~\eqref{eq:czz} after the ramp has crossed the critical point.
In the exact solution, slower ramps leave the chain closer to the ferromagnetic ground state, and the correlations spread to longer distances as the ramp time grows~\cite{Cincio2007,Kolodrubetz2012}, as shown in Fig.~\ref{fig:1d_corr}.
However, DTWA does not reproduce this.
We observe that its correlations grow only slightly with the ramp time and remain short-ranged, decaying within a few lattice spacings even for the slowest ramps.
DTWA therefore cannot capture how the correlations at different distances grow with the ramp time.

We observe the same behavior in the correlation length [Fig.~\ref{fig:fig2-correlation-length}(a)].
The exact correlation length approaches the Kibble--Zurek power law until it reaches half the system size, where it plateaus, whereas the DTWA correlation length grows only for the fastest ramps and then saturates after a few lattice spacings, independent of the ramp time and of the system size.
DTWA also produces a state that no quantum system can reach.
For slow ramps, its nearest-neighbor correlator rises above one [Fig.~\ref{fig:fig2-correlation-length}(b)], violating the quantum bound, while the exact correlator approaches one from below.

We evaluate the kink density from two angles [Fig.~\ref{fig:fig2-correlation-length}(c)].
The first uses the nearest-neighbor correlator through Eq.~\eqref{eq:nkink}, as for the exact solution.
The second counts, in each trajectory, the bonds whose two spins have opposite signs of $s^z_i$, which treats every classical spin as pointing up or down.
The exact kink density decreases with the ramp time as predicted by Kibble--Zurek.
DTWA cannot capture this from either angle.
Because its nearest-neighbor correlator exceeds one for slow ramps, the first estimate becomes negative, while the second stops decreasing and saturates.
DTWA agrees with the exact kink density only for the fastest ramps.

\begin{figure}[t]
  \centering
  \includegraphics[width=0.8\columnwidth]{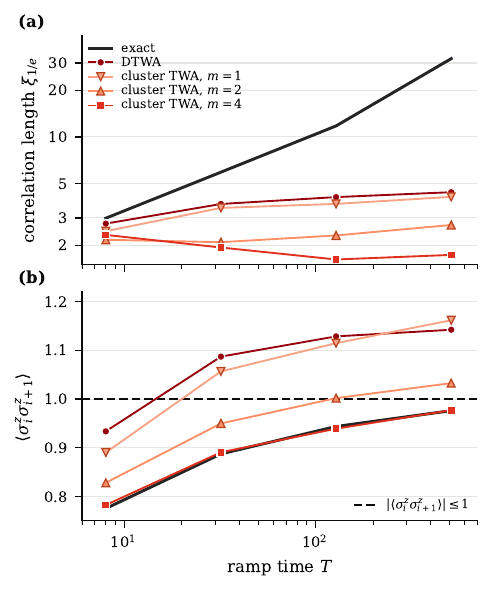}
  \caption{Cluster TWA on the chain at the end of the ramp, $s=1$, versus ramp time $T$ for $N=64$, compared with the exact solution (black line) and DTWA, for cluster sizes $m=1$, 2 and 4. (a) Correlation length $\xi_{1/e}$. (b) Nearest-neighbor correlator, with the quantum bound $|\langle\sz_i\sz_{i+1}\rangle|\le 1$ dashed; the exact curve lies beneath $m=4$. Larger clusters restore the nearest-neighbor correlator but shorten the correlation length rather than lengthening it.}
  \label{fig:fig4-cluster-twa}
\end{figure}

Treating clusters of spins exactly does not remove this limitation (Fig.~\ref{fig:fig4-cluster-twa}).
With larger clusters, the nearest-neighbor correlator approaches the exact result and returns below the quantum bound [Fig.~\ref{fig:fig4-cluster-twa}(b)], but the correlation length becomes shorter rather than longer [Fig.~\ref{fig:fig4-cluster-twa}(a)].
The improvement of the nearest-neighbor correlator, however, comes from a cancellation. Bonds inside a cluster, which are treated exactly, become too weak, whereas bonds that cross a cluster boundary still exceed the quantum bound, and their average lies close to the exact value. Correlations also do not pass through an exactly treated cluster, so the correlations between clusters decay faster as the clusters grow, which shortens the correlation length.
Cluster TWA therefore improves the averaged nearest-neighbor correlator but cannot capture the long-distance correlations that carry the Kibble--Zurek scaling.

On the chain, we also use the fast-ramp protocol of Eq.~\eqref{eq:schmitt} to calibrate the scaling collapse of Sec.~\ref{sec:2d3d}.
Applied to the exact solution, the collapse recovers an exponent close to the Kibble--Zurek value, and DTWA gives a similar exponent, so on the chain the difference between DTWA and the exact solution appears mainly after the critical point.

These results are consistent with the earlier finding that DTWA misses long-distance correlations in quenches of the chain~\cite{Czischek2018}, and they extend it to ramps and Kibble--Zurek scaling.
We identify their cause in Sec.~\ref{sec:mechanism}.

\section{All-to-all coupling}
\label{sec:lmg}
In this section, we turn to all-to-all coupling, the opposite limit to the chain.
Every spin interacts equally with all others, the linear ramp crosses the transition at $J_c=h$, that is at $s_c=1/2$, and the exact dynamics in the Dicke subspace is available for large systems.
We compare the order parameter $m_z^2$ of Eq.~\eqref{eq:m2} along the whole ramp.

\begin{figure}[t]
  \centering
  \includegraphics[width=0.88\columnwidth]{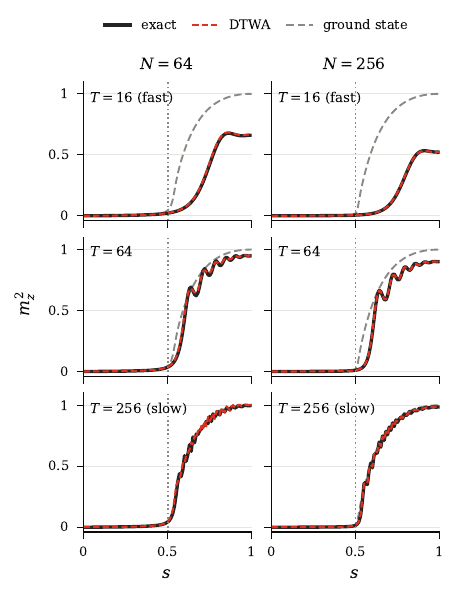}
  \caption{Order parameter $m_z^2$ along the linear ramp with all-to-all coupling, for $N=64$ (left) and $N=256$ (right) and ramp times $T=16$, 64 and 256 (top to bottom). Black lines show the exact solution in the Dicke subspace, red dashed lines show DTWA, and gray dashed lines show the exact ground state. The dotted line marks the critical point $s_c=1/2$. The oscillations after the critical point are coherent oscillations of the collective spin, present in the exact solution and reproduced by DTWA.}
  \label{fig:fig5-all-to-all}
\end{figure}

The exact order parameter stays close to zero until the critical point and then grows (Fig.~\ref{fig:fig5-all-to-all}).
Slow ramps follow the ground state closely, while fast ramps lag behind it and end the ramp less ordered.
After the critical point, the order parameter oscillates, because the ramp leaves the collective spin slightly excited and it oscillates about the new ordered state.
DTWA reproduces these curves for every ramp time, including the delayed growth for fast ramps, the oscillations, and the value reached at the end of the ramp.
Corrections to the mean-field limit are of order $1/N$, so we expect the agreement to improve further for larger systems.

The delayed growth for fast ramps is physical and not an artifact of DTWA.
In both the exact solution and DTWA, the order grows from quantum fluctuations of size $1/\sqrt N$ in the initial state, so its onset after the critical point is delayed, and the delay grows slowly with $N$.
DTWA captures this because it samples these initial fluctuations correctly.

With all-to-all coupling, the dynamics is that of a single large collective spin, whose quantum fluctuations are small and whose classical motion is regular.
The square and cubic lattices of the next section lie between these two limits.

\section{Two and three dimensions}
\label{sec:2d3d}
In this section, we consider the square and cubic lattices, where each spin has four and six neighbors, respectively.
Their critical points are known from QMC, $h_c/J=3.04438(2)$ and $5.15813(6)$~\cite{BloteDeng2002}, so the linear ramp crosses them at $s_c=0.247$ and $s_c=0.162$.
We ask whether DTWA orders, whether it places the transition correctly, and whether it gives the right Kibble--Zurek exponent, and we treat these questions in turn.

\subsection{Ordering and critical points}
\begin{figure}[t]
  \centering
  \includegraphics[width=0.88\columnwidth]{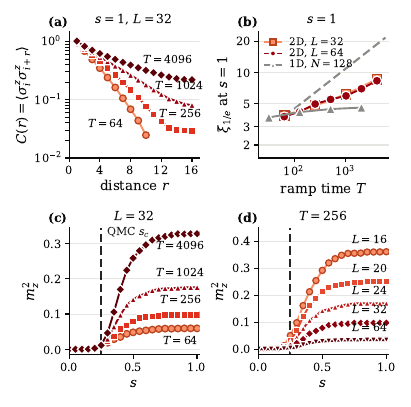}
  \caption{DTWA on the square lattice. (a) Correlation function $C(r)$ at the end of the ramp, $s=1$, for $L=32$ and ramp times $T=64$, 256, 1024 and 4096; points smaller than twice their statistical error are omitted. (b) Correlation length $\xi_{1/e}$ at $s=1$ versus ramp time for $L=32$ and 64, compared with DTWA on the chain for $N=128$ (gray triangles). The dashed line is the Kibble--Zurek power law $T^{0.39}$ expected on the square lattice, anchored at the first point for $L=64$. (c) Order parameter $m_z^2$ along the ramp for $L=32$ and the same ramp times as in (a). (d) Order parameter along the ramp for $T=256$ and $L=16$ to 64. The vertical dashed lines mark the QMC critical point $s_c=0.247$. The correlations grow with the ramp time, but the order parameter decreases with the system size.}
  \label{fig:fig9-2d-summary}
\end{figure}

On the square lattice, the DTWA correlations at the end of the ramp extend further as the ramp slows down [Fig.~\ref{fig:fig9-2d-summary}(a)], and the correlation length keeps growing with the ramp time without depending on the system size [Fig.~\ref{fig:fig9-2d-summary}(b)].
This differs from the chain, where the correlation length saturates after a few lattice spacings, although the growth stays well below the Kibble--Zurek power law.

We also see that the order parameter grows after the critical point, and it grows more for slower ramps [Fig.~\ref{fig:fig9-2d-summary}(c)].
At a fixed ramp time, however, it decreases steadily with the system size [Fig.~\ref{fig:fig9-2d-summary}(d)], so DTWA does not reach long-range order on the square lattice within the ramp times we can access.

\begin{figure}[t]
  \centering
  \includegraphics[width=0.8\columnwidth]{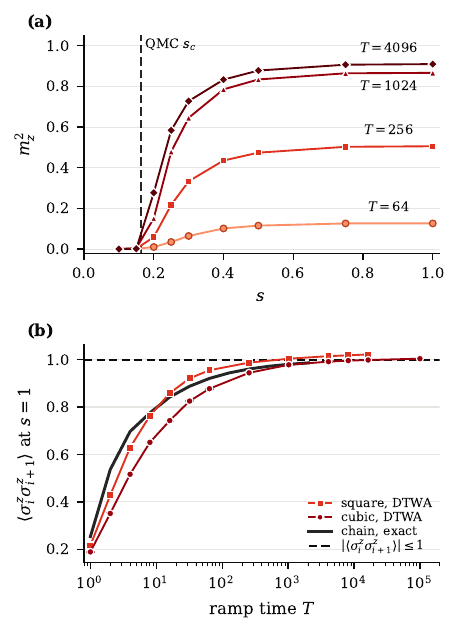}
  \caption{(a) Order parameter $m_z^2$ along the linear ramp on the cubic lattice from DTWA, for $L=16$ and ramp times $T=64$, 256, 1024 and 4096. The dashed line marks the QMC critical point $s_c=0.162$. Slow ramps leave the lattice close to full order. (b) Nearest-neighbor correlator at the end of the ramp, $s=1$, versus ramp time, from DTWA on the square lattice ($L=32$) and the cubic lattice ($L=16$). The black line shows the exact result on the chain for reference, and the dashed line marks the quantum bound $|\langle\sz_i\sz_{i+1}\rangle|\le 1$. DTWA exceeds the bound on both lattices, on the cubic lattice only for the slowest ramp, by about four standard errors (error bars).}
  \label{fig:cubic-bound}
\end{figure}

The cubic lattice behaves differently.
There, the order parameter rises sharply just after the QMC critical point and grows strongly with the ramp time [Fig.~\ref{fig:cubic-bound}(a)].
Fast ramps leave the lattice only weakly ordered, whereas for the slowest ramps the order parameter approaches its maximum, and the curves for the two slowest ramps nearly coincide, which indicates that the order has saturated.

On both lattices, DTWA also violates the quantum bound, as on the chain [Fig.~\ref{fig:cubic-bound}(b)].
On the square lattice, the nearest-neighbor correlator exceeds one for slow ramps, and on the cubic lattice only for the slowest ramp we simulate.
The violation therefore sets in at slower ramps and becomes weaker as the number of neighbors grows, and on the cubic lattice it is small.

\begin{figure}[t]
  \centering
  \includegraphics[width=\columnwidth]{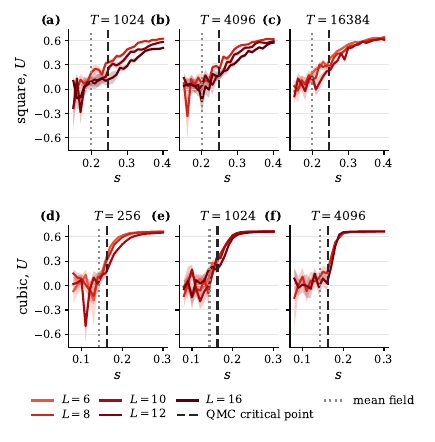}
  \caption{Binder cumulant $U$ along the linear ramp from DTWA, on the square lattice (top row) for $T=1024$ and 4096 with $L=8$, 12 and 16, and for $T=16384$ with $L=6$, 8 and 10, and on the cubic lattice (bottom row) for $T=256$ with $L=6$, 8 and 10, for $T=1024$ with $L=6$ to 12, and for $T=4096$ with $L=8$ and 12. Shaded bands show one standard error. Dashed lines mark the QMC critical point and dotted lines the mean-field critical point $h_c=zJ$, with $z$ the number of neighbors. On the cubic lattice, the curves rise together just after the QMC critical point and reach the ordered value $2/3$ for every system size.}
\label{fig:fig11-binder-fss}
\end{figure}

To locate the transition that DTWA produces, we use the Binder cumulant of Eq.~\eqref{eq:binder}, evaluated along the ramp for several system sizes at a fixed ramp time (Fig.~\ref{fig:fig11-binder-fss}).
The cumulant stays close to zero in the disordered phase and rises toward $2/3$ as the system orders, and the point where the curves for two system sizes cross marks the transition.
Close to zero the curves are dominated by statistical noise, so we use only crossings at intermediate values of the cumulant.

On the cubic lattice, the curves for different system sizes rise steeply together just after the QMC critical point [Fig.~\ref{fig:fig11-binder-fss}(d)--(f)]. Their crossings lie at the QMC critical point for intermediate ramps, but for the slowest ramp they move to the ordered side.
After the transition, the cumulant reaches its ordered value $2/3$ for every system size, which is consistent with long-range order, in contrast with the square lattice.
DTWA therefore orders close to, but somewhat after, the quantum critical point, and clearly later in the ramp than the mean-field prediction.

On the square lattice, the crossings have not converged within the system sizes and ramp times we can reach, and the curves remain noisy [Fig.~\ref{fig:fig11-binder-fss}(a)--(c)].
Still, the cumulant starts to rise around the QMC critical point rather than at the mean-field value, so DTWA indicates where the transition lies on the square lattice as well, although our data do not establish long-range order there.

\subsection{Kibble--Zurek exponents}
\begin{figure}[t]
  \centering
  \includegraphics[width=0.8\columnwidth]{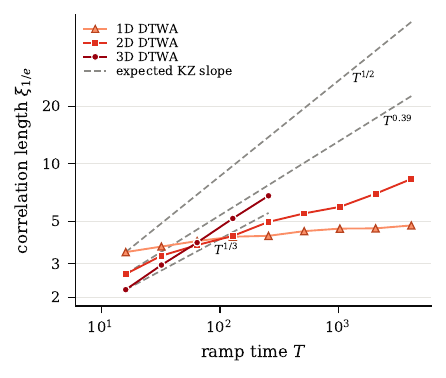}
  \caption{Correlation length $\xi_{1/e}$ at the end of the ramp versus ramp time, from DTWA on the chain ($N=128$), the square lattice ($L=64$) and the cubic lattice ($L=32$). Dashed lines show the Kibble--Zurek power laws expected in each dimension, $T^{1/2}$, $T^{0.39}$ and $T^{1/3}$, anchored at the first DTWA point. The exact correlation length on the chain follows $T^{1/2}$ [Fig.~\ref{fig:fig2-correlation-length}(a)]. Only points with $\xi_{1/e}<L/2$ are shown. For slower ramps on the cubic lattice the correlations stay above $e^{-1}$ across the whole box.}
  \label{fig:fig10-dimensions}
\end{figure}

We now ask how the DTWA correlations grow with the ramp time. Figure~\ref{fig:fig10-dimensions} shows the correlation length at the end of the ramp on the chain and on the square and cubic lattices, together with the Kibble--Zurek power laws of Table~\ref{tab:kz}. On the chain, the DTWA correlation length barely grows and stays at a few lattice spacings, as in Fig.~\ref{fig:fig2-correlation-length}(a). On the square lattice, it keeps growing over the whole range of ramp times, but more slowly than the Kibble--Zurek power law. On the cubic lattice, it grows faster than the power law.

In two and three dimensions, the correlation length at the end of the ramp is not set by the frozen correlation length $\hat\xi$ of Eq.~\eqref{eq:kz} alone. After the critical point has been crossed, ordered domains continue to grow by coarsening, so the correlation length at the end of the ramp can grow faster with the ramp time than $\hat\xi$~\cite{Chandran2012,Schmitt2022}, which may explain the fast growth on the cubic lattice. The fast-ramp protocol of Eq.~\eqref{eq:schmitt} avoids this, because it stops at the critical point, where we test the scaling form of Eq.~\eqref{eq:collapse}.

\begin{figure}[t]
  \centering
  \includegraphics[width=0.88\columnwidth]{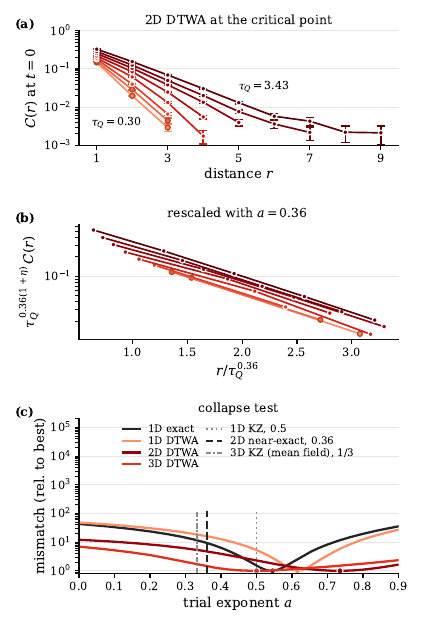}
  \caption{Scaling collapse at the critical point for the fast-ramp protocol of Eq.~\eqref{eq:schmitt}. (a) Correlation function $C(r)$ at $t=0$ from DTWA on the square lattice ($L=48$) for quench times $\tau_Q=0.30$ to $3.43$, from light to dark. Error bars show one standard error, and only points with $C(r)$ larger than twice its standard error are shown. (b) The same correlations rescaled according to Eq.~\eqref{eq:collapse} with the near-exact exponent $a=0.36$. The curves do not collapse. (c) Mismatch between the rescaled curves as a function of the trial exponent $a$, relative to its minimum, for the exact solution and DTWA on the chain ($N=128$) and for DTWA on the square ($L=48$) and cubic ($L=32$) lattices. Correlations below $0.03$ are left out, and below $0.003$ on the cubic lattice, where the correlations are shorter. Dots mark the best exponents, whose precise values depend on this cutoff. The dashed line marks the near-exact value $0.36$ on the square lattice, the dotted line the Kibble--Zurek value $1/2$ on the chain, and the dash-dotted line the Kibble--Zurek value $1/3$ on the cubic lattice.}
  \label{fig:fig12-schmitt-comparison}
\end{figure}

On the square lattice, the DTWA correlations at the critical point decay more slowly as the quench time increases [Fig.~\ref{fig:fig12-schmitt-comparison}(a)], but rescaling them with the near-exact exponent $0.36$ does not bring them onto one curve [Fig.~\ref{fig:fig12-schmitt-comparison}(b)]. To find the best collapse, we vary the exponent $a$ in $\hat\xi=\tau_Q^{\,a}$ of Eq.~\eqref{eq:collapse} and measure how far apart the rescaled curves lie on their common range [Fig.~\ref{fig:fig12-schmitt-comparison}(c)]. The same test recovers an exponent close to the Kibble--Zurek value from the exact solution on the chain (Sec.~\ref{sec:1d}), and Ref.~\cite{Schmitt2022} found a collapse in the same window of quench times on the square lattice. For DTWA, the best exponent lies well above the Kibble--Zurek value on both the square and the cubic lattice [Fig.~\ref{fig:fig12-schmitt-comparison}(c)]. Its precise value depends on the smallest correlation included in the comparison, but it stays well above $0.36$ on the square lattice and above $1/3$ on the cubic lattice for every choice.

Taken together, DTWA does not reproduce the Kibble--Zurek growth of the correlations. In the fast-ramp protocol on the square and cubic lattices, its correlations grow faster with the quench time than predicted, whereas at the end of slow ramps they grow more slowly on the square lattice and stop growing on the chain, so no range of ramp times shows the Kibble--Zurek scaling on the chain or the square lattice. On the cubic lattice, the correlations grow faster than predicted both at the critical point and at the end of the ramp, where coarsening also contributes. DTWA thus builds order close to the transition but cannot capture its universal dynamics.

\section{Mechanism}
\label{sec:mechanism}
In the previous sections, we found that with short-range coupling DTWA builds order close to the transition on the square and cubic lattices but cannot capture its Kibble--Zurek dynamics. On the chain, its correlation length saturates at a few lattice spacings, independent of the system size, and its nearest-neighbor correlations exceed the quantum bound. In this section, we show that these behaviors are consistent with a common origin. DTWA evolves classical spins, and the quantum fluctuations of the initial state enter only through random initial conditions, which act like heat. With short-range coupling, the classical spins therefore relax toward a thermal state at an effective temperature. We test this picture by comparing DTWA with this classical thermal state, first on the chain and then on the square and cubic lattices.
\begin{figure}[t]
  \centering
  \includegraphics[width=0.88\columnwidth]{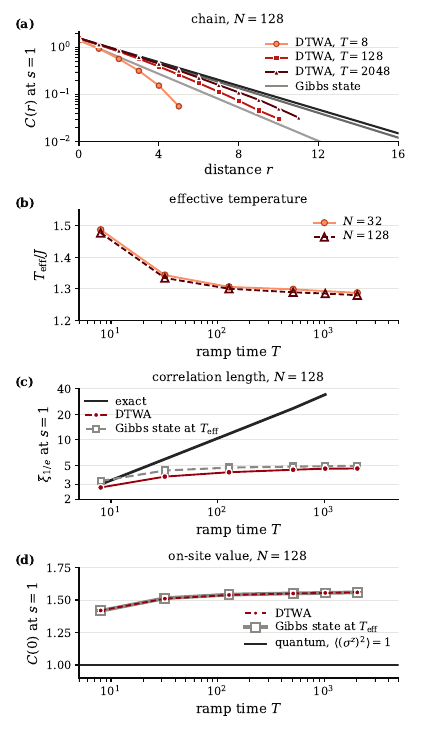}
  \caption{Classical thermal state on the chain at the end of the ramp, $s=1$. (a) Correlation function $C(r)$ from DTWA (red) and from the classical Gibbs state with the same nearest-neighbor correlation $C(1)$ (gray), for $N=128$ and ramp times $T=8$, 128 and 2048, from light to dark. Only points with $C(r)$ larger than twice its standard error are shown. (b) Effective temperature $T_{\rm eff}$ from this fit versus ramp time, for $N=32$ and 128. (c) Correlation length versus ramp time from the exact solution, from DTWA and from the Gibbs state at $T_{\rm eff}$, for $N=128$. (d) On-site value $C(0)$ versus ramp time from DTWA and from the Gibbs state at $T_{\rm eff}$, for $N=128$. The horizontal line marks the quantum value one.}
  \label{fig:mechanism}
\end{figure}

In DTWA, every sampled spin has length $\sqrt3$, which the dynamics conserves, and the spins evolve under the Weyl symbol $H_{\rm W}$ of the Hamiltonian. If this classical dynamics thermalizes, the state is described by the Gibbs state of $H_{\rm W}$ at an effective temperature $T_{\rm eff}$,
\begin{equation}
\begin{gathered}
  H_{\rm W}=-h\sum_i s^x_i - J\sum_{\langle ij\rangle}s^z_is^z_j,\qquad |\mathbf s_i|=\sqrt3,\\
  \rho_{\rm Gibbs}\propto e^{-H_{\rm W}/T_{\rm eff}},\qquad C_{\rm Gibbs}(1;T_{\rm eff})\equiv C_{\DTWA}(1),
\end{gathered}
\label{eq:gibbs}
\end{equation}
We fix $T_{\rm eff}$ by a single condition, that the Gibbs state reproduces the nearest-neighbor correlation $C(1)$ of DTWA. Every other quantity of the Gibbs state is then a prediction that can be compared with DTWA.

We first test this at the end of the ramp, $s=1$, where the field vanishes and only the $z$ components enter $H_{\rm W}$. Since every $s^z_i$ is then conserved, the state at the end of the ramp is not formed there but records the state reached while the field was still finite, and it freezes as the field goes to zero. The $z$ component of a spin distributed uniformly on a sphere of radius $\sqrt3$ is itself distributed uniformly on $[-\sqrt3,\sqrt3]$, so on the chain the Gibbs correlations follow exactly from a transfer matrix,
\begin{equation}
\begin{gathered}
  C_{\rm Gibbs}(r)=\frac{\mathrm{Tr}\big[\hat s^z \mathcal{K}^{r} \hat s^z \mathcal{K}^{N-r}\big]}{\mathrm{Tr}\,\mathcal{K}^{N}}\quad (s=1),\\
  \mathcal{K}(s^z,s^{z\prime})=e^{Js^zs^{z\prime}/T_{\rm eff}},\qquad s^z\in[-\sqrt3,\sqrt3]\ \text{uniform}.
\end{gathered}
\label{eq:transfer}
\end{equation}
The on-site value $C(0)=\overline{(s^z)^2}$ is then a prediction. It agrees with DTWA for every system size and ramp time we simulated, and it lies well above the quantum value of one [Fig.~\ref{fig:mechanism}(d)]. At short distances the Gibbs correlations also agree closely with DTWA, whereas at long distances the DTWA correlations lie below them, with a gap that narrows as the ramp slows down [Fig.~\ref{fig:mechanism}(a)]. DTWA therefore approaches the thermal state from below, and in all our runs the correlation length of the Gibbs state lies above that of DTWA.

Since $T_{\rm eff}$ is fitted separately at each ramp time, we can follow how it changes as the ramp slows down [Fig.~\ref{fig:mechanism}(b)]. The fitted temperature drops for the fastest ramps and then decreases only slowly, staying at a finite value that is the same for all system sizes, so slower ramps do not cool the DTWA state. In the quantum system, by contrast, slower ramps leave the state closer to the ground state, so its effective temperature would approach zero.

This finite temperature sets a ceiling on the correlations. A chain with short-range coupling cannot order at any finite temperature, so the correlation length of the classical Gibbs state stays finite. The DTWA correlation length stays just below this thermal value for all ramp times, whereas the exact correlation length keeps growing [Fig.~\ref{fig:mechanism}(c)]. This picture is consistent with the saturation of the DTWA correlation length, its independence of the system size and the absence of Kibble--Zurek scaling (Sec.~\ref{sec:1d}). Since at $h=0$ the Gibbs correlation length on the chain is fixed by $C(1)$, this consistency is not an independent test, and the independent tests are $C(0)$ and the shape of $C(r)$ discussed above. The same picture also accounts for $C(1)$ exceeding the quantum bound. A classical spin of length $\sqrt3$ is not restricted to $|s^z|\le1$, and in the thermal state the coupling tilts the spins toward the $z$ axis, so neighboring spins can be correlated by more than one.

\begin{figure}[t]
  \centering
  \includegraphics[width=0.88\columnwidth]{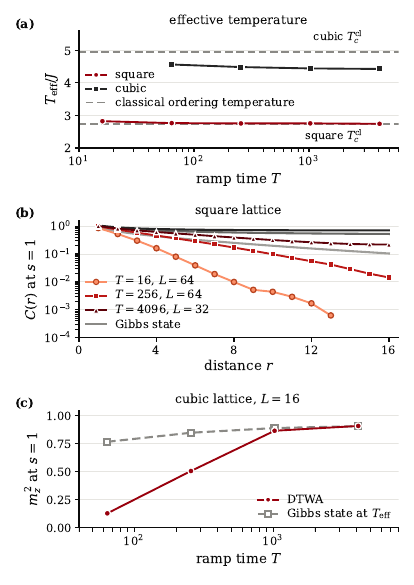}
  \caption{Classical thermal state on the square and cubic lattices at the end of the ramp, $s=1$. (a) Effective temperature $T_{\rm eff}$, fitted to the nearest-neighbor correlation $C(1)$ of DTWA, versus ramp time on the square lattice ($L=64$, and $L=32$ for $T=4096$) and on the cubic lattice ($L=16$). Dashed lines mark the ordering temperatures of the classical models, obtained from crossings of the Binder cumulant. (b) Correlation function $C(r)$ on the square lattice from DTWA (red) and from the Gibbs state at $T_{\rm eff}$ (gray), from light to dark for increasing ramp time. (c) Order parameter $m_z^2$ on the cubic lattice from DTWA and from the Gibbs state at $T_{\rm eff}$, for $L=16$.}
  \label{fig:mechanism2d}
\end{figure}

On the square and cubic lattices, we sample the Gibbs state by classical Monte Carlo. The fitted temperature again barely changes with the ramp time [Fig.~\ref{fig:mechanism2d}(a)]. Unlike the chain, however, these classical models order below a finite temperature, and where $T_{\rm eff}$ lies relative to it decides how DTWA orders. On the square lattice, $T_{\rm eff}$ lies at the classical ordering temperature, so DTWA ends near a classical critical point. That $T_{\rm eff}$ stays at this temperature for every ramp time is striking, and we have no explanation for it. It orders only partly, and its long-range correlations stay well below the Gibbs values [Fig.~\ref{fig:mechanism2d}(b)]. On the cubic lattice, $T_{\rm eff}$ lies below the classical ordering temperature, and for slow ramps DTWA reaches the ordered Gibbs state [Fig.~\ref{fig:mechanism2d}(c)]. The order that DTWA builds in three dimensions, found in Sec.~\ref{sec:2d3d}, is thus consistent with classical thermal order.

This picture also suggests why with short-range coupling the DTWA correlations grow with the ramp time at the wrong rate. The DTWA state does not freeze out through the quantum Kibble--Zurek mechanism. Instead, it relaxes like a classical magnet at the temperature $T_{\rm eff}$, so there is no reason for its correlations to follow the quantum exponents.

\section{Discussion and outlook}
\label{sec:discussion}
Taken together, our results suggest that DTWA becomes more reliable near the transition as each spin interacts with more partners (Table~\ref{tab:summary}), although on the lattices studied here the number of neighbors and the dimension change together.
On the chain, with two neighbors, DTWA neither orders nor captures the growth of the correlations.
On the square and cubic lattices, with four and six neighbors, it builds order, which sets in close to the QMC critical point, but its correlations do not grow with the Kibble--Zurek exponents.
With all-to-all coupling, it follows the exact order parameter closely.
This trend is consistent with earlier studies of DTWA~\cite{Kunimi2021,Muleady2023}.

\begin{table}
  \caption{Summary of DTWA along the linear ramp, compared with exact or near-exact results.}
  \label{tab:summary}
  \begin{ruledtabular}
  \small\setlength{\tabcolsep}{2pt}
  \begin{tabular}{lcccc}
    Geometry & Neighbors & Order & Onset of order & Kibble--Zurek \\
    \hline
    Chain & 2 & none & none & no \\
    Square & 4 & partial & near QMC & no \\
    Cubic & 6 & strong & after QMC & no \\
    All-to-all & $N-1$ & as exact & as exact & not tested \\
  \end{tabular}
  \end{ruledtabular}
\end{table}

A similar trend appears in quantum-inspired annealing, whose exponents approach the quantum ones as the dimension grows~\cite{Xie2026}.
Mean-field theory, however, contains no quantum fluctuations, whereas in DTWA it is precisely these fluctuations, entering through the random initial conditions, that act as heat and limit the accuracy (Sec.~\ref{sec:mechanism}).

For applications of DTWA near quantum critical points, our results suggest a simple guide.
In two and three dimensions, DTWA can indicate roughly where a transition lies, but it should not be trusted for its precise location or its critical exponents.
A nearest-neighbor correlator above one is a warning sign that DTWA has entered the classical thermal regime.

Several questions remain open.
The effective temperature $T_{\rm eff}$ is fitted rather than derived from first principles, and it has not been tracked along the ramp, so it is not known whether the transition found by DTWA lies where $T_{\rm eff}$ crosses the classical ordering temperature.
A quantitative theory of the growth of the DTWA correlations, which would predict the exponents found in Sec.~\ref{sec:2d3d}, is also missing, and the classical chaos that drives the thermalization has not been demonstrated directly.
Power-law couplings $1/r^\alpha$ on the chain would allow a continuous scan between the short-range chain and all-to-all coupling, and a comparison with tensor-network simulations in two dimensions would test DTWA beyond the critical point.
DTWA has also been applied near a transition in an open, dissipative transverse-field Ising model, where it produced classical rather than quantum critical exponents~\cite{SinghWeimer2022}.
This is another case of DTWA behaving classically, although in that open system classical exponents may also be the physically correct answer.
DTWA has also been used to extract critical scaling at a nonequilibrium squeezing transition in bilayer spin models with power-law couplings, where the transition is reached by time evolution rather than by a ramp through a ground-state critical point~\cite{Duha2025,Duha2026}. There, the exponents agree across several lattice geometries, and an independent Bogoliubov analysis confirms the phase boundary~\cite{Duha2026}. Comparing the two settings may help to clarify when DTWA captures critical scaling.

\begin{acknowledgments}
We thank Identity Quantum Computing for providing the computing hardware and for funding this work.
Parts of the simulation and analysis code, and the editing of the manuscript, were carried out with the assistance of an AI model (Claude, Anthropic). The authors designed the study, checked all code, results and text, and take full responsibility for the content of this article.
The simulation code, analysis scripts and data are available on GitHub~\cite{CodeRepo} and Zenodo~\cite{CodeZenodo}.
\end{acknowledgments}

\appendix
\section{Exact and near-exact references}
\label{app:exact}
We compare DTWA with exact results wherever they exist, which is the case for the chain and for all-to-all coupling. On square and cubic lattices, only the QMC critical points and, in two dimensions, near-exact Kibble--Zurek results are available.

\subsection{Chain: free fermions}
On the chain, the Jordan--Wigner transformation maps the TFIM onto free fermions~\cite{Pfeuty1970}. We first exchange the spin axes and then represent the rotated spins by fermions,
\begin{equation}
  \tau^z=\sx,\quad \tau^x=\sz,\qquad \tau^z_j=1-2c^\dagger_jc_j,
  \label{eq:jw}
\end{equation}
so that the initial state $|{+}x\rangle^{\otimes N}$ becomes the fermion vacuum. The Hamiltonian conserves the fermion parity, and the ramp therefore stays in the even-parity (Neveu--Schwarz) sector, whose momenta are $k=\pi(2n-1)/N$ with $n=1,\dots,N/2$ for even $N$. Each pair of momenta $(k,-k)$ then evolves independently as a two-level system with the Hamiltonian
\begin{equation}
  H_k=\begin{pmatrix}0 & 2iJ\sin k\\ -2iJ\sin k & 4(h-J\cos k)\end{pmatrix}
  \ \text{on}\ \{|0\rangle,\ c^\dagger_kc^\dagger_{-k}|0\rangle\}.
  \label{eq:hk}
\end{equation}
We integrate it with the exponential midpoint rule, propagating every pair in each time step with the exact exponential of $H_k$ evaluated at the middle of the step.

In the rotated frame $\sz_j=\tau^x_j$, so the correlation function $C(r)$ becomes a string of fermion operators. We write it in terms of the Majorana operators $a_j=c_j+c_j^\dagger$ and $b_j=i(c_j^\dagger-c_j)$, and Wick's theorem turns it into the Pfaffian
\begin{equation}
\begin{split}
  \langle\sz_0\sz_r\rangle &= \mathrm{Pf}\,K,\qquad K_{ab}=\mathrm{Im}\,\langle x_ax_b\rangle,\\
  (x_1,\dots,x_{2r}) &= (b_0,a_1,b_1,\dots,b_{r-1},a_r),
\end{split}
\label{eq:pf}
\end{equation}
which we evaluate with the Parlett--Reid algorithm. This expression holds for ground states and for time-evolved states alike. We validate the solution against exact diagonalization in App.~\ref{app:validation}. As a further check, it reproduces the Kibble--Zurek scaling of the kink density, Eq.~\eqref{eq:kzkink}.

\subsection{All-to-all coupling: Dicke subspace}
For all-to-all coupling, the Hamiltonian of Eq.~\eqref{eq:lmg} conserves the total spin, and the initial state lies in the symmetric sector $S=N/2$. We therefore evolve the state numerically exactly in this $(N+1)$-dimensional Dicke subspace, in the eigenbasis of $S^z$, where the Hamiltonian is tridiagonal. We integrate the Schr\"odinger equation with a fourth-order Runge--Kutta (RK4) integrator, with a time step small enough that halving it leaves the results unchanged. Because RK4 does not conserve the norm exactly, we normalize the state before each measurement.

\subsection{Square and cubic lattices}
On square and cubic lattices no exact dynamics is available. For the location of the transition we use the QMC critical points of Ref.~\cite{BloteDeng2002}, given in Sec.~\ref{sec:2d3d}. For the dynamics, we compare in two dimensions with the near-exact Kibble--Zurek results of Ref.~\cite{Schmitt2022}, obtained with the fast-ramp protocol of Eq.~\eqref{eq:schmitt}.

\section{Validation of the numerical codes}
\label{app:validation}
We first checked the free-fermion code against exact diagonalization of a short chain, for ground states and for a quench, and found agreement to numerical precision. For the ramps we use free fermions rather than exact diagonalization, because a Runge--Kutta integration of the full state vector with a fixed time step becomes inaccurate at strong coupling, where the norm of the Hamiltonian grows.

For DTWA, the limit $h=0$ provides an analytic test. Each $s^z_j$ is then conserved, and spin $i$ precesses about the $z$ axis at the constant rate $2J\sum_{j\in\partial i}s^z_j$. Averaging $s^x_i(t)$ over the independent random signs of Eq.~\eqref{eq:sampling} gives $\langle\sx\rangle=\cos^{q}(2Jt)$ for a spin with $q$ neighbors, which coincides with the exact quantum result. The same argument gives $\langle\sx\rangle=\cos^{N-1}(2Jt/N)$ for all-to-all coupling. DTWA is therefore exact at $h=0$, and our codes reproduce these results within statistical error.

Away from this limit DTWA is an approximation. After a quench on a short chain, the DTWA nearest-neighbor correlator follows the exact result at early times and departs from it later. This deviation does not change when the time step is reduced, and it is much larger than the statistical error, so it stems from the truncation and not from the time integration.

For cluster TWA, a single cluster that contains the whole ring has no mean-field coupling. Equation~\eqref{eq:ctwa_eom} is then linear in $M_c$, and since the average of $M_c(0)$ is the initial density matrix, the average of $M_c(t)$ is the exact density matrix at all times.

\section{Simulation parameters}
\label{app:params}
Table~\ref{tab:runs} lists the system sizes, time steps and numbers of trajectories of all DTWA and cluster TWA data sets. We checked that halving the time step leaves the results unchanged within statistical error. Statistical errors are standard errors over independent batches of trajectories, with jackknife errors for the Binder cumulant of Eq.~\eqref{eq:binder}, and every observable is averaged over all sites and lattice axes. The simulation code, analysis scripts and data behind every figure are available online~\cite{CodeRepo,CodeZenodo}.

\begin{table}
  \caption{System sizes, time steps $\delta t$ and numbers of trajectories of the DTWA and cluster TWA data sets.}
  \label{tab:runs}
  \begin{ruledtabular}
  \begin{tabular}{llcc}
   Data set & Sizes & $\delta t$ & Trajectories \\ \hline
   Chain, linear ramps & $N=32$--$256$ & 0.01 & 480 \\
   Chain, fast ramps & $N=128$ & 0.002 & 960 \\
   Chain, cluster TWA & $N=64$ & 0.02 & 96--192 \\
   All-to-all & $N=64,\,256$ & 0.02 & 960--3840 \\
   Square, linear ramps & $L=6$--$64$ & 0.02 & 48--768 \\
   Square, fast ramps & $L=48$ & 0.002 & 480 \\
   Cubic, linear ramps & $L=6$--$32$ & 0.02 & 16--240 \\
   Cubic, fast ramps & $L=32$ & 0.002 & 960 \\
  \end{tabular}
  \end{ruledtabular}
\end{table}

\end{document}